\documentclass{ws-procs975x65}

\def\be{\begin{eqnarray}}
\def\ee{\end{eqnarray}}

\def\beq{\begin{equation}}
\def\eeq{\end{equation}}

\begin{document}

\title{Instability of strong magnetic field and neutrino magnetic dipole moment}
\author{Hyun Kyu Lee$^*$ }

\address{Department of Physics, Hanyang University University\\
Seoul, 04763, Korea\\
$^*$E-mail: hyunkyu@hanyang.ac.kr}



\begin{abstract}

Vacuum instability of the strong electromagnetic field  has been discussed since long time ago. The instability of the strong electric field  due to creation of electron pairs is one of the examples, which is known as Schwinger process.   What matters are  the  coupling of particles to the electromagnetic field and the mass of the particle to be produced.   The critical electric field for electrons in the minimal coupling  is $E_c \sim \frac{m^2}{e} $.  Spin 1/2 neutral particles but with magnetic dipole moments  can interact with the electromagnetic field through   Pauli coupling. The instability of the particular vacuum under the strong magnetic field  can be formulated  as the emergence of imaginary parts of the effective potential. In this talk, the development of the imaginary part in the effective potential as a function of the magnetic field strength is discussed for the configurations of the uniform magnetic field and the inhomogeneous magnetic field.
Neutrinos are the lightest particle(if not photon or gluon) in the ``standard model", of which  electromagnetic property is poorly known experimentally.  Recently  the observation of neutrino oscillation shows the necessity of neutrino masses. It implies that the standard model is subjected to be modified such that non-trivial electromagnetic structure of neutrino should be reconsidered  although they are assigned to be neutral. And the possibility of anomalous  electromagnetic form factor is an open question theoretically and experimentally.
In this talk,  the implication of non-vanishing magnetic dipole moment of neutrinos is also  discussed:  the instability of the strong magnetic field and the enhancement of neutrino production in high energy collider experiments.

\end{abstract}

\bodymatter

\section{Introduction}

There are various indications of the ultrastrong magnetic fields , $B > B_c \sim 10^{13}$ G, from astrophysical observations and terrestrial accelerator experiments and there have been continuous efforts in finding their definite evidence in Nature.  Among the examples considered so far are  strong magnetic fields captured on   magnetar and  gamma ray bursts(GRB) central engine and also created in noncentral heavy-ion collisions.
Magnetars\cite{magnetar} are considered to be  neutron stars, where  strong magnetic fields of typically $10^{13} \sim  10^{15}$ G are the main source of energy.  Even the stronger field strength is expected inside magnetars.  One of the viable models of GRB central engine is to make use the idea of tapping the rotational energy of black hole by the strong magnetic field\cite{LBW}. It is estimated to be $\sim 10^{15}$  G particularly   for long and energetic  bursts.
In noncentral heavy-ion collisions, strong magnetic fields can also be created by  the two electric currents in opposite directions generated by  two colliding nuclei. it is expected that the magnetic fields in RHIC
Au + Au collisions and LHC Pb + Pb collisions  can be as large as $10^{19}$ G and $10^{20}$ G respectively\cite{Huang}. These field strengths  are  much  stronger than  the critical magnetic field and  offer an interesting  opportunity to study the effect of super-strong electromagnetic fields beyond the classical electrodynamics and beyond the standard model of electro-weak interaction.  One of the interesting questions with given  such strong electromagnetic fields is the stability of the field configuration  or equivalently  whether there is any instability, which leads to  the particle production.   Vacuum instability of the strong electromagnetic field  has been discussed since long time ago. The instability of the strong electric field  due to the creation of electron pairs is one of the examples, which is known as Schwinger process\cite{schwinger, Kim}.   What matters are  the  coupling of particles to the electromagnetic field and the mass of the particle to be produced.   The critical electric field for electron  is $E_c \sim \frac{m^2}{e} $.  If it is possible to imagine a charged particle lighter than the electron, the critical field can be lowered down to the field strength produced in the future high intensity laser  and will be subjected to be tested in the laboratory.

The pair production of fermions in a purely magnetic field configuration is shown to be absent\cite{dunnehall}. Therefore, the pair production of minimally interacting particles is considered to be a purely electric effect. However, while the minimal coupling derived by the local gauge invariance is of fundamental
nature, there appear also non-minimal couplings as well in the form of the effective theory. Pauli introduced a non-minimal coupling of spin-1/2 particles with electromagnetic fields, which can be interpreted as an effective interaction of fermions with an anomalous magnetic moment\cite{Pauli}.  Hence    for the neutral fermions, which have no minimal coupling to electromagnetic fields,  the nonvanishing magnetic moments may be  the primary  window through which the electromagnetic interaction of neutral fermions  can be probed with the Pauli interaction.

It is known the spatial inhomogeneity of the magnetic field exerts force on  the magnetic dipole moment through the Pauli interaction. It  plays a similar role analogous to the electric field for the creation of  charged particle pairs with the minimal coupling. The possibility of pair production of the neutral fermions in a purely magnetic field configuration with spatial inhomogeneity has been demonstrated  in 2+1 dimension\cite{lin}. The production
rate in 3+1 dimension has been calculated explicitly for the magnetic fields with a spatial inhomogeneity\cite{LY1, Gitman}, which can be approximated as
\be
\varpi  \propto m^4 e^{-a m^2/|\mu B'|}
\ee
 analogues to the Schwinger process.

The instability of the particular vacuum under the strong magnetic field  can be formulated  as the emergence of imaginary parts of the effective potential. For uniform magnetic fields which interact with spin-1/2 fermions through the Pauli interaction\cite{LY2}, it is found that the non-vanishing imaginary part develops for a magnetic field stronger than the  critical field $B_c$, whose strength is the ratio of the fermion mass to its magnetic moment,
$B_c =  \frac{m}{\mu}$,

\be
 {\cal I}(V_{eff}) = \frac{m^4}
{48 \pi} ( \frac{|\mu B|}{m} -1)^3
( \frac{|\mu B|}{m} +3).
\ee
In section 2, the calculations of the effective potential and vacuum decay rates for the neutral fermion with Pauli coupling to electromagnetic field will be  reviewed.

The implication of non-vanishing magnetic dipole moment of neutrinos,   the instability of the strong magnetic field and the  neutrino production through the Pauli coupling in high energy collider experiment,   is  discussed  in section 3.

\section{Imaginary part of effective potential and pair production}

The instability of the electro-weak vacuum for a strong magnetic field was discussed long time ago\cite{AHN}.  The one-loop effective potential considering the   weak-boson($W$)  loop is found to have imaginary part under the pure magnetic background,
\be
{\cal I}(V_{eff}) = \frac{e^2}{8\pi} B^2(1-\frac{m^2_W}{eB})\Theta(B - \frac{m_W^2}{e^2}) \label{ImVW}
\ee
where $m_W$ is the mass of $W$-boson . In the limit of $m_W \rightarrow 0$, eq.(\ref{ImVW})  agrees with  the effective potential given by Nielson and Olesen \cite{NO}.  It was argued that the instability could be avoided if  the condensation of $W$ and $Z$ bosons\cite{AHN} appears with the strong magnetic field.  The basic reason for the emergence of an imaginary part for $B > \frac{m_W^2}{e^2}$ is that the energy eigenvalue crosses zero at $B =\frac{m_W^2}{e^2}$ because of the anomalous magnetic moment of $W$ boson.

One can see that there is also a level crossing of energy eigenvalue of a neutrino   with Pauli interaction for a strong enough magnetic field. For a uniform magnetic field, the energy eigenvalues of
the Hamiltonian are given by
\be
E = \pm \sqrt{p^2_l + (\sqrt{m^2 + p^2_t} \pm \mu B)^2} \label{Epauli}
\ee
where $p_l$ and $p_t$ are respectively the longitudinal and the transversal momentum to the magnetic field direction. One can see that for a magnetic field stronger than the critical field $Bc = m/\mu$ , the ground state with $p_l = p_t = 0$ crosses the zero energy state.  This indicates the possible instability of the magnetic field configuration beyond critical field strength as in electro-weak instability \cite{LY3}.  For a uniform magnetic field,  the imaginary part of the one-loop effective action is calculated explicitly\cite{LY2}
\be
{\cal I}(V_{eff}) = \frac{m^4}{48 \pi} ( \frac{|\mu B|}{m} -1)^3
( \frac{|\mu B|}{m} +3)\Theta(|\mu B| - m).
\ee
which takes the similar form as in eq. (\ref{ImVW}).  It is interesting to note that the development of the imaginary part associated with a level crossing has been also demonstrated in the different contexts\cite{GF}. The state occupied in the negative energy sea becomes a particle state and the vacant positive energy state plunges into the negative sea to make an antiparticle state.  The instability can be interpreted as the pair creation at the expense of  the magnetic field strength.  Then the  particle production rate
is given approximately by
\be
\varpi  \sim 2 {\cal I}(V_{eff}) =  \frac{m^4}
{24 \pi} ( \frac{|\mu B|}{m} -1)^3
( \frac{|\mu B|}{m} +3)\Theta(|\mu B| - m).\label{rate}
\ee

For a  nonuniform magnetic field,  the inhomogeneity of the magnetic field coupled directly to the magnetic dipole moment plays an interesting role analogous to the electric field for a charged particle: The non-zero gradient of the magnetic field can exert a force on a magnetic dipole moment.  Then the vacuum production of neutral fermions with a non-zero magnetic moment in an inhomogeneous magnetic field is possible.  As a simple example,  a static magnetic field configuration
 with a constant gradient, $B'= dB_z/dx$, along x-direction,
\be
B_z(x) = B_0 + B'x.
\ee
has been considered.
It is not necessary to consider an infinitely extended ever-increasing magnetic field to meet the linear magnetic field configuration. Because the particle production rate density is a local quantity, it is sufficient to have a uniform gradient magnetic field in the Compton wavelength scale of the particle.  The imaginary part of the one-loop effective potential is calculated in the integral form\cite{LY2},
\be
{\cal I} (V_{eff}) &=& -\frac {m^4}{4\pi^2\lambda \kappa} \int^\infty_0 \frac{dv}
{v^2} [\sqrt{v \coth v}F(\frac{\lambda}{\kappa} \tanh v, \lambda v)  -\frac{1}{2} \cos \lambda v - \frac{ \lambda v}{ 12\kappa} \sin \lambda v] \nonumber \\
 & & - \frac{m^4}{ 4 \pi^2 \lambda^2 } \int^\infty_0 \frac{dv}{v^3}[(v \coth v)^{3/2}-  1 - \frac{v^2}{2}] \sin \lambda v,
\ee where
\be
v &=& s \mu B', ~~ \lambda = \frac{m^2}{|\mu B'|}, ~~ \kappa = \frac{m^2}{|\mu B|^2}, \\
F(a,b) &=& \int^1_0 d\xi (1-\xi)\cos(a\xi^2-b)
\ee
Numerically it is found that it can be fitted to  an  analytic expression\cite{LY2},
\be
{\cal I}(V_{eff})   \propto m^4 e^{- b \frac{m^2}{|\mu B'|}},
\ee
which is decreasing exponentially with respect to the inverse of the field gradient. The qualitative feature of this result can  be compared   with  the expression obtained recently by Gavrilov and Gitman\cite{Gitman}.

\section{Neutrino magnetic dipole moment and its implications}

Neutrinos are the lightest particles(if not photon or gluon) in the ``standard model", of which  electromagnetic property is poorly known experimentally.  Moreover  the observation of neutrino oscillation shows the necessity of neutrino masses. It implies that the standard model is subjected to be modified such that non-trivial electromagnetic structure of neutrino should be reconsidered  although they are assigned to be neutral.
 While there is no observation which is not consistent with the electric-charge neutrality of neutrinos, the possibility of anomalous  electromagnetic form factor is an open question theoretically and experimentally\cite{GS}.

 So far there is no experimental evidence of the magnetic dipole moment of the neutrinos but one can not simply rule out the magnetic moment just because of the electrical neutrality. For example, in a minimal extension of the standard
model to incorporate the neutrino mass the anomalous magnetic moment of a  neutrino is known to be developed  in one loop calculation \cite{fujikawa}, $\mu_\nu = \frac{3eG_F}{8\sqrt{2}\pi^2} m_\nu $. Here one can notice that the non-zero mass is essential to get a non-vanishing magnetic moment.  There are strong evidences from the neutrino oscillation observations that neutrinos  have nonzero-masses \cite{numass}.   However,    the simple extensions of the standard model can give only  much smaller  magnetic moment, $ \sim  10^{-20}\mu_B$ if we take the neutrino mass as 1 eV ,  than the experimental bound.  GEMMA collaboration\cite{gemma} observed    that   the neutrino magnetic moment is bounded from above by the value $\mu_\nu < 2.9 \times  10^{-11} \mu_B $ and the bound obtained from Borexino\cite{borex} data  gave $\mu_\nu \leq  3.1 \times  10^{-11}\mu_B$ .   Since a wide range of neutrino magnetic moments is possible up to the current laboratory upper limit beyond standard model,  it is interesting to investigate  the phenomenological consequences of the nonvanishing magnetic moment up to the current bounds, which may constrain the bound tighter than the direct measurements  of the neutrino magnetic dipole  moment.

For the vacuum instability, the typical field strength(or critical field strength) is given by $B_c = m_\nu/\mu_{\nu}$.   Taking the possible magnetic moment to be as large as the experimental upper bound $\mu_\nu = 10^{-11}\mu_B$ and the mass of the neutrino $m_\nu \sim 10^{-2} eV$ constrained by the neutrino oscillations, the critical field strength is estimated to be $B_c \sim  10^{17}$ G, which is not far from the field strength inferred on the magnetar surface and might be possible inside the magnetar. With vacuum instability, the particle production rate per unit time per unit volume is typically of order $m_\nu^4$ for the critical field strength. As a possible environment, let us consider the pair creation
of neutrinos with non-zero magnetic dipole moments  from  the very strongly
magnetized compact objects with $B = B_c$ and radius $R_B$. With $R^3_B$
as an effective volume of the magnetosphere, the production rate can be estimated\cite{LY4} as
 \be
 \varpi_\nu \sim 10^{39}(\frac{m_\nu}{10^{-2}eV} )^4(\frac{B}{B_c} )^4(\frac{R_B}{10km})^3 / s.
 \ee
 Using the currently estimated masses of neutrinos the corresponding luminosity   is estimated to be  $L \sim 10^{25}$ erg/s  assuming  $E_\nu \sim m_\nu$,  which      is   much weaker than the X-ray luminosity of magnetars, $L_X \sim 10^{35}$erg/s. However, since the time scale is much longer than the Schwinger process by a factor of $(m_e/m_\nu)^4$, it  can be a continuous source of neutrinos.  A larger luminosity for observation can be easily obtained with a massive particle but the magnetic moment should be increasing to keep the critical field strength, which significantly constrains models for particles involved.  For the vacuum instability beyond critical field strength in case of the uniform magnetic field, the interpretation of vacuum instability as the neutrino pair production may not be the whole story.  It is because there may be a possibility  that  the instability could be avoided if  a nontrivial  condensation\cite{AHN} develops along with the critical  magnetic field, which is subjected to further discussion.

The pair production of neutrinos in collider experiments can be  enhanced   if endowed with the nonvanishing magnetic dipole moment. The pair production of  neutrinos are expected  from the annihilation of charged particles in colliding experiments  through photon channel\cite{klpy} in addition to the weak boson channel in the standard model.
One of the interesting features of Pauli coupling is  that the angular distribution of pair creation  peaks at $\theta = \pi/2$.  It is compared to the angular
distribution of the standard model process, minimum at $\theta = \pi/2$ but
maximum for $\theta =  0$ and $\pi$. For the Majorana neutrinos,
the  pairs with  different flavors  can be  produced due to the transit magnetic moment couplings. The the angular distribution peaks at $\theta=\pi/2$ with respect to the beam direction, which is the enhancement of the neutrino production at right angle in contrast to weak Z-channel.  It is interesting to note that the flavor mixing angle can be inferred if the flavor correlation in pairs can be  measured in the future experiments\cite{L}. The detectors located around the right angle to the beam direction can measure the back-to-back correlations
in the pair production, where the production rate is supposed to be maximum. Most of the detectors are using electromagnetic triggers at the end stations, which implies corresponding charged leptons($l_\alpha$ and $l_\beta$), for example  electrons or
muons, are those to be detected finally. The Majorana pairs
in mass eigenstates, $\nu_1$ and $\nu_2$, are produced through the Pauli coupling and interact
weakly with other particles to produce charged leptons to be detected.
To simplify the situation let us consider  only two mass eigenstates which are mixed states  of weak eigenstates, $\nu_\alpha$ $\nu_\beta$,
\be
\nu_1 = \cos\delta ~ \nu_\alpha + \sin\delta ~ \nu_\beta,  ~~ \nu_2 = -\sin\delta ~ \nu_\alpha + \cos\delta ~ \nu_\beta
\ee
where
$\delta$  is the  mixing angle.  Now consider two targets A and B
placed at the opposite side of the center of mass at the
right angle to the beam direction.  At the targets the neutrinos can produce the corresponding charged leptons via the charged current weak interaction(W-boson).   The back-to-back correlation, R, defined by the
product of the detection rates of $\alpha$ type leptons at the target A and $\beta$ type leptons  at the target B, can be given by
\be
R = \frac{1}{2}( 1 + \cos^2 2\delta).
\ee
Since the correlation $R$ turns  out to be  dependent only on the mixing angle, $\delta$,
it may serve as a clean measurement of mixing angles in the collider experiments. For example  the maximum value, $R= 1$, is obtained
for the cases of no-mixing($\delta=0$), $R=1/2$ for the maximal mixing($\delta=\pi/4$).

These features are quite different aspect from standard model, which can be  clearly  distinguished in the high energy collider experiment\cite{L}.
Sine the strength of Pauli coupling to the photon is determined by the neutrino magnetic dipole moment, which is very small, the pair creation cross section through photon channel by the Pauli coupling  is smaller that that of weak boson($Z$) channel until the center of mass energy of the colliding particle is sufficiently high.  To get some idea of the energy scale for which
the Pauli  coupling becomes  dominant, we can define the energy scale, $E_{0.1}$, for which the total cross section becomes $10 \%$ of the standard model,
\be
E_{0.1} = 10^3(\frac{10^{-11} \mu_B}{\mu_\nu}) TeV.
\ee  It is a bit higher than LHC energy, $E_{CM} \sim 10$ TeV and also higher than the GZK cut-off energy of ultra high energy cosmic rays, $E_{CM} \sim 100$ TeV.  So to see the distinguished feature  from standard model, the higher energy  collider experiments than present LHC seem to be  needed.

The magnetic field is expected  be created by  the two electric currents in opposite directions generated by  two colliding nuclei. It turns out the field strength may be very large as $10^{19} - 10^{20}$G at RHIC and LHC.  It  gives $10^4$ times stronger Pauli coupling than expected at the surface of the  magnetars .  The  magnetic vacuum instability due to magnetic dipole moment of neutrino can be  much more enhanced. How to identify experimental observables corresponding to the vacuum instability and particle productions  that are sensitive to the neutrino magnetic dipole moment may be  an interesting and challenging subject to be discussed.

\section{Acknowledgments}

The author would like to thank Dmitry Gitman,  Sang Pyo Kim and Yongsung Yoon for useful comments and discussions.


\bodymatter

\end{document}